# Elastic impact of a sphere with an elastic half-space: numerical modeling and comparison with experiment

I.A. Lyashenko[1,2], E. Willert[1], V.L. Popov[1,3,4]
[1]Berlin University of Technology, 10623 Berlin, Germany
[2]Sumy State University, 40007 Sumy, Ukraine
[3]National Research Tomsk State University, 634050 Tomsk, Russia
[4]National Research Tomsk Polytechnic University, 634050 Tomsk, Russia
E-mail: v.popov@tu-berlin.de

Numerical simulations of the dynamics of an elastic collision between a rigid sphere and an elastic half-space are carried out. We assume an Amontons-Coulomb frictional force with a fixed coefficient of friction between the contacting surfaces during the impact. As a result of modeling a dimensionless function, describing the tangential restitution, is found. It depends only on a small set of governing dimensionless parameters and allows the determination of the sphere's tangential velocity and angular velocity after the collision. This function is used to calculate the tangential velocity recovery rate and the cyclic rotation frequency of a sphere for the parameters of a particular experiment on particle reflection of an aluminum alloy from a glass plate. The obtained results with high accuracy coincide with the experimental data, which confirms the adequacy of the proposed numerical model.

## 1 INTRODUCTION

In many technological applications it is important to understand the processes during collisions of solid particles. The collision characteristics determine, for example, the dynamics of granular media [1], [2], [3], [4]. Hitherto, many issues remain unresolved in this area, as the precise dynamics of the collision – including partial slip, full slip or no-slip in the contact area during various phases of the impact – can be very complicated [5].

A classical theory for the case without slip i.e. an infinite coefficient of friction, based only on conservation laws and the rolling condition, can be found in textbooks on mechanics [6]. It is, however, contradictory, since the body is considered to be elastic, but, at the same time, the kinematic condition of a rigid-body rotation for the sphere is used. The tangential contact, as well as micro-slip in the contact zone, was first described in [5] (MBF theory). The authors of this work made use of the theory for the normal contact of elastic spheres, proposed by Hertz in 1882 [7], and Mindlin's theory for the respective tangential contact problem [8]. Later, Barber proposed an analytical description [9], but only for collision phases without slippage. The MBF theory was confirmed experimentally by the authors themselves in [10], and later also by other collectives [11], [12], [13]. A review of existing collision models and their experimental confirmation can be found in the book [14].

In recent works, Popov and others have shown that the Hertz-Mindlin theory can be exactly reproduced by replacing the real three-dimensional contact [15] with a contact between a certain modified plain profile and a linearly elastic foundation of independent springs [16], [17]. This approach has been called the Method of Dimensional Reduction (MDR) and results in a great simplification of the analysis of dynamic contact problems. Thus, the collision of elastic spheres without slip or adhesion could rigorously be analyzed in [18]. The generalizations of a collision with a finite friction coefficient in the contact zone (without adhesion) as well as of an adhesive collision without slip have been carried out in further recent papers [19], [20]. In connection with the fact that a large amount of experimental material has been accumulated on the dynamics of collisions, we had the idea to describe a real experiment using a MDR model. This is the subject of the present work. It shows that the simulation of the collision of a spherical body with a half-space coincides almost perfectly with the experimental results, which confirms the validity of the models developed earlier.

## 2 PROBLEM DEFINITION

Consider a collision of an elastic ball with mass $m$ and radius $R$ with an elastic half-space, as shown in Fig. 1.

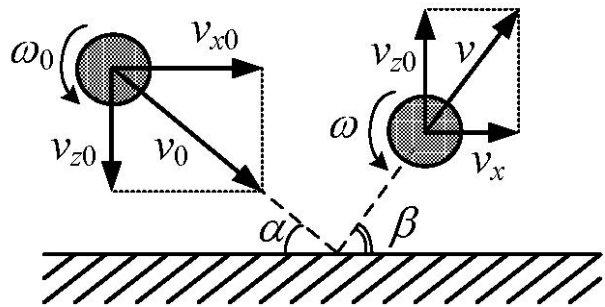

Fig. 1. Schematic representation of the collision of a spherical body with a half-space at an angle to the surface. The normal velocity component does not change after collision in absolute magnitude, since an elastic collision is considered.

Let the elastic moduli of the ball and half-space be $E_1$ and $E_2$; the Poisson ratios shall be $v_1$ and $v_2$, the shearing moduli $G_1$ and $G_2$ respectively and the material densities of the ball and half-space shall be denoted as $\rho_1$ and $\rho_2$. The initial velocity of the ball $v_0$ is decomposed into a vertical $v_{z0}$ and horizontal $v_{x0}$ component before the collision event, the initial cyclic rotation frequency of the ball is $\omega_0$. A part of the kinetic energy is transferred to the body with which the ball contacted after the collision event, and as a result the velocities after the collision are $v$, $-v_{z0}$, $v_x$, $\omega$. If all velocities are known, it is easy to determine the slip angle for the falling body $\alpha°$ and also the slip angle $\beta°$ of the ball rebounding from the surface.

First, we give the classical solution of this problem without slip. For a homogeneous sphere (the moment of inertia being $I = (2/5)mR^2$), the classical solution [6]

$$\bar{v}_x = \frac{5}{7}v_{x0} - \frac{2}{7}R\omega_0, \tag{1}$$

$$\bar{\omega} = \frac{2}{7}\omega_0 - \frac{5}{7}\frac{v_{x0}}{R}, \tag{2}$$

can be obtained from simple balance equations for the momentum and angular momentum and the rolling condition

$$v_x + \omega R = 0. \tag{3}$$

The change of kinetic energy due to the collision will be

$$\Delta E = \frac{m}{2}\left(\bar{v}_x^2 - v_{x0}^2\right) + \frac{I}{2}\left(\bar{\omega}^2 - \omega_0^2\right) = -\frac{m}{7}(v_{x0} + R\omega_0)^2. \tag{4}$$

It can be seen from expression (4) that the impact is not absolutely elastic, since the energy is not conserved (its change is negative). Note that the solutions (1) and (2) do not contain material parameters of the shear modulus type, but depend only on the initial conditions. This will occur only for a narrow range of parameters and in most practical cases will be incorrect, because the kinematic condition (3) cannot be valid throughout the whole contact and the assumption of its validity at the moment of detachment from the surface is arbitrary and unjustified. In fact, condition (3) can only be correct at one time during the contact because of the elastic properties of the ball. The behavior of the system is much more complicated than the description given by the classical solution.

# 3 SIMULATION OF THE ELASTIC COLLISION OF A SPHERICAL BODY WITH AN ELASTIC HALF-SPACE WITH A FINITE COEFFICIENT OF FRICTION IN THE CONTACT ZONE

The comprehensive simulations of the collision of an elastic spherical body with an elastic half-space with a fixed coefficient of friction between the contacting surfaces have been carried out in [19]. Here we give a brief description of the simulation procedure. Note, that we restrict ourselves on the contact between a rigid sphere and an effective elastic half-space with the effective moduli

$$\frac{1}{E^*} = \frac{1-v_1^2}{E_1} + \frac{1-v_2^2}{E_2}, \tag{5}$$

$$\frac{1}{G^*} = \frac{2-v_1}{4G_1} + \frac{2-v_2}{4G_2}. \tag{6}$$

Nevertheless, the more general problem exhibits no additional features. The normal and tangential displacements of the center of gravity of the spherical body are denoted as $u_z$ (down-directed) and $u_x$ (directed to the right along the motion), the normal and tangential forces as $F_N$ (direction coincides with the direction of $u_z$) and $F_f$ (direction coincides with $u_x$). Then, the equations of motion for the sphere are written in the form

$$m\ddot{u}_z = -F_N, \tag{7}$$

$$m\ddot{u}_x = -F_f, \tag{8}$$

$$I\ddot{\varphi} = -F_f R, \tag{9}$$

where $\varphi$ is the angle of rotation of the sphere, the change of which determines the angular velocity through the relation $\dot{\varphi} = \omega$. To model the normal and tangential contact in the collision, we will use the Method of Dimensionality Reduction (MDR) [17]. Within MDR, the following steps need to be done: First, the three-dimensional elastic half-space is replaced by a linearly elastic foundation consisting of an array of independent springs with a distance $\Delta x$ between them. For each spring, the values of tangential and normal stiffness are calculated as

$$\Delta k_z = E^* \Delta x, \tag{10}$$

$$\Delta k_x = G^* \Delta x. \tag{11}$$

In the next step, the axisymmetric three-dimensional profile $z = f(r)$ is transformed into a flat profile $g(x)$ according to the rule

$$g(x) = |x| \int_0^{|x|} \frac{f'(r)}{\sqrt{x^2 - r^2}} \, dr. \tag{12}$$

Finally, the transformed flat profile is brought into contact with the elastic foundation and moved according to the desired (enforced or resulting) protocol. It was shown in [17], that the relations of the MDR model between the contact forces, displacements and contact radii reproduce the three-dimensional solution. The solution of the contact problem within the framework of the MDR has the same accuracy as the solution of Cattaneo [21] and Mindlin [8]: there is an error in the case of an arbitrary value of Poisson's ratio; however, it is insignificant [22].

For a sphere of radius $R$ the profile in the vicinity of the contact has the form

$$f(r) = \frac{r^2}{2R}.$$

According to (12) this function corresponds to a one-dimensional profile

$$g(x) = \frac{x^2}{R}. \tag{13}$$

If the vertical displacement of the center of mass $u_z$ is measured from the first moment of contact, it will coincide with the depth of indentation; the vertical displacement of a spring in the elastic base inside the contact area at the point $x_i$ is defined as

$$\tilde{u}_z(x_i) = u_z - g(x_i). \tag{14}$$

The contact radius $a$ is determined by the condition

$$\tilde{u}_z(a) = u_z - g(a) = 0. \tag{15}$$

To solve the tangential contact problem first the increment

$$d\tilde{u}_x(x_i) = du_x + R d\varphi \tag{16}$$

is added at each time step to the horizontal displacement of the springs. After that the no-slip condition (resulting from a local Coulomb law for the spring forces)

$$\mu E^* \tilde{u}_z(x_i) \geq |G^* \tilde{u}_x(x_i)|. \tag{17}$$

is checked for all springs at each iteration to account for local slip. For all slipping springs, violating the condition (17), a new displacement value is set,

$$\tilde{u}_x(x_i) = \pm \frac{\mu E^* \tilde{u}_z(x_i)}{G^*}, \tag{18}$$

where the sign coincides with the sign $\tilde{u}_x(x_i)$ in expression (17), i.e. before assigning the new value. After that, the normal and tangential forces are calculated by summation

$$F_N = E^* \Delta x \sum_{cont} \tilde{u}_z(x_i), \quad (19)$$

$$F_f = G^* \Delta x \sum_{cont} \tilde{u}_x(x_i), \quad (20)$$

Since $F_N$ and $F_f$ are now determined, we can numerically solve the system of equations (7) – (9).

In the course of a numerical experiment, the function $P(\gamma, \alpha)$ is defined in [19]; its values allow us to find the velocity and the cyclic frequency of the ball after the collision if the analogous quantities are known prior to the impact. This function is written as follows:

$$P(\gamma, \alpha) = \frac{7}{2} \frac{v_x - \bar{v}_x}{v_{x0} + R\omega_0} = \frac{7}{5} \frac{R(\omega - \bar{\omega})}{v_{x0} + R\omega_0}, \quad (21)$$

where the coefficient of friction $\mu$ enters (21) through the parameter $\alpha$:

$$\alpha(\mu) = \frac{\mu v_{z0}}{|v_{x0} + R\omega_0|}. \quad (22)$$

In (21) we also introduced the function

$$\gamma = \sqrt{\frac{7}{2} \frac{G^*}{E^*}}, \quad (23)$$

which takes into account the material characteristics of the contacting bodies. Note, that the functions $P(\gamma, \alpha)$ and $\alpha(\mu)$ are not defined for $(v_{x0} + R\omega_0) = 0$. In this case we have pure normal contact, and therefore it is obvious, that $v_x = v_{x0}$ and $\omega = \omega_0$. The change in the kinetic energy of the colliding sphere due to the contact processes is given by

$$\frac{7 \Delta E}{m(v_{x0} + R\omega_0)^2} = -1 + (P(\gamma, \alpha))^2. \quad (24)$$

When the collision process was modeled, equations (7) – (9) were solved numerically by the explicit Euler method. Fig. 2 shows the dependence obtained in the result of the numerical experiment, where the results of numerous computer experiments fall on a single surface, which makes it universal.

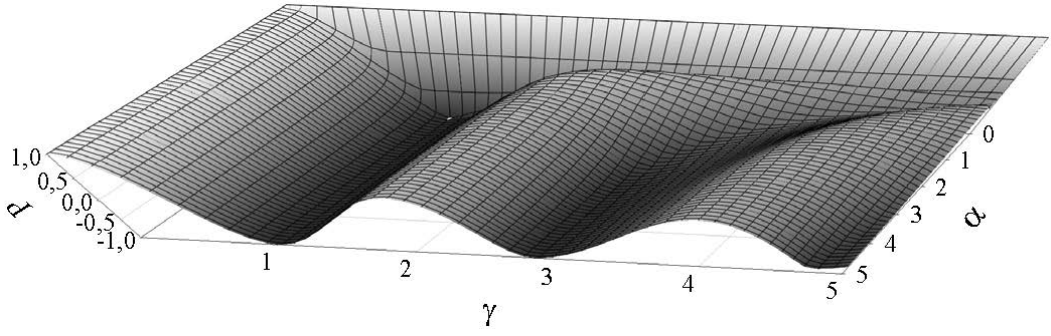

Fig. 2. Numerically calculated dependence $P(\gamma,\alpha)$ (21) as a function of the parameters $\gamma$ (23) and $\alpha$ (22). The resulting surface is the result of a large number of numerical experiments with a different set of initial conditions, radii, elastic moduli, and the mass of a spherical body.

The calculated function $P(\gamma,\alpha)$ (21) completely determines the behavior of the system. This function is shown over a wide range of the parameter $\gamma$ (23). If the contacting bodies are made of the same material or either one of them is rigid, then $G^*/E^* = 2(1-\nu)/(2-\nu)$. Due to thermodynamic stability Poisson's ratio can only take values $-1 < \nu \leq 1/2$ [23]. Thus, the resulting inequalities $2/3 < G^*/E^* < 4/3$ lead to the following range of values of the dimensionless parameter $\gamma$ for isotropic bodies:

$$1.52 < \gamma < 2.16. \qquad (25)$$

During the contact of the ball with the half-space, different friction modes with the presence of slippage and adhesion are possible. These modes are set by two critical values of the parameter $\alpha$ [19]:

$$\alpha_{c1} = \frac{2\gamma^2}{7(2\gamma^2-1)}, \qquad (26)$$

$$\alpha_{c2} = \frac{2\gamma^2}{7}. \qquad (27)$$

Conditionally, friction modes in a collision can be divided into three different scenarios (I, II, III). These modes are shown in the diagram in Fig. 3. When the inequality $\alpha \geq \alpha_{c2}$ (field I) is satisfied, no-slip prevails in the contact during the compression stage of the impact. Partial slippage (field II) occurs if $\alpha_{c1} < \alpha < \alpha_{c2}$. And the last field III (total slippage) is observed for $\alpha \leq \alpha_{c1}$. We note that in the case of $\gamma = 1$ both critical quantities coincide $(\alpha_{c1} = \alpha_{c2})$, therefore, the above is valid only for $\gamma > 1$, when $\alpha_{c2} > \alpha_{c1}$ (which corresponds to the actual region (25)). The areas I, II and III discussed above are shown in the diagram in Fig. 3. Note, that for $\gamma = 1$ and $\alpha > \alpha_c = 2/7$ the absolute value of $P(\gamma,\alpha)$ would be one, i.e. the impact would be completely conserving the kinetic energy. Within the range (25) this, however, is impossible.

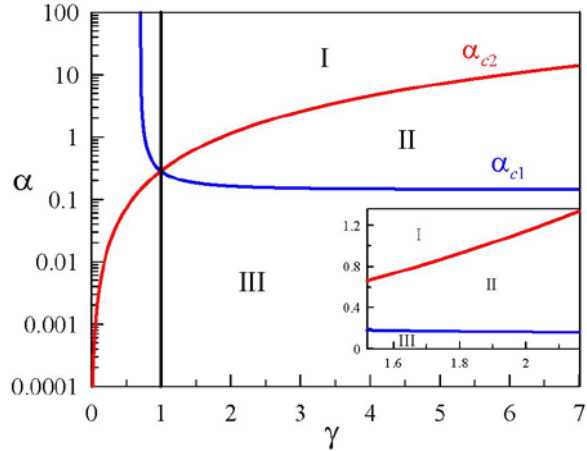

Fig. 3. Diagram of collision modes with a finite friction coefficient. The actual area (25) is shown in the inset to the figure.

Fig. 4a shows the dependence $P(\gamma,\alpha)$ as a function of the parameter $\alpha$ (22) for 17 values of the parameter $\gamma$ (23) from the actual range (25).

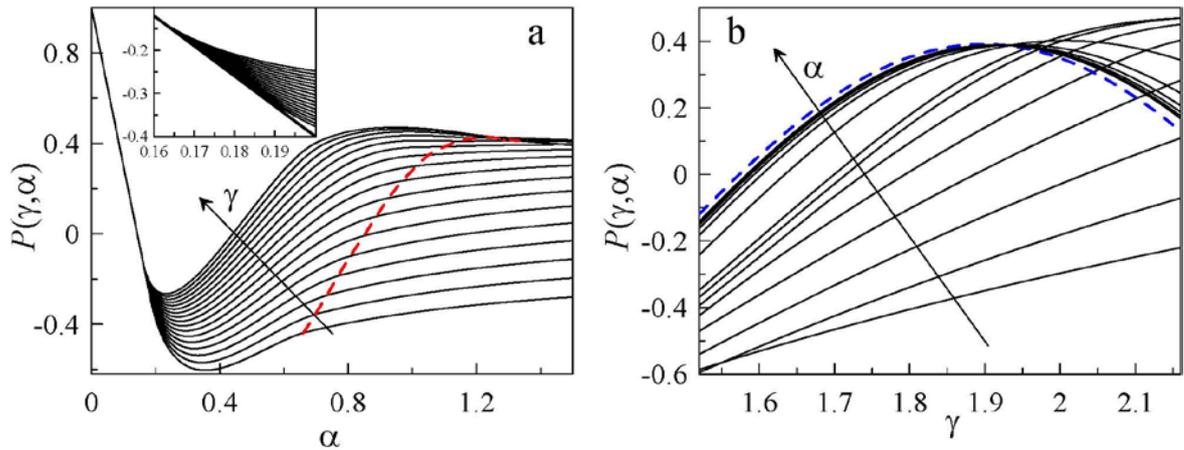

Fig. 4. (a) The dependencies $P(\gamma,\alpha)$ as a function of the parameter $\alpha$ (22) for fixed values $\gamma$ (23). The value $\gamma$ varies from 1.52 to 2.16 with an interval 0.04. Increasing direction $\gamma$ shown by the arrow. The dashed line shows the critical value $\alpha_{c2}$ (27). The solid line in the inset to the figure illustrates the dependence (28); (b) dependence of $P(\gamma,\alpha)$ as a function of the parameter $\gamma$ (23) for fixed values $\alpha$ (22). The lower 8 curves are constructed with increasing $\alpha$ from 0.3 to 1.0 in increments of 0.1. Further, $\alpha$ increases from 1 to 12 in increments of 1.0. In the figure, the magnification direction $\alpha$ is indicated by an arrow. The dashed line shows the dependence corresponding to complete sticking in the contact zone [18].

The figure shows, that a universal behavior, independent of the value $\gamma$, is observed for all curves for small $\alpha$. Here, all curves lie on a single straight line given by the equation [19]

$$P(\gamma,\alpha) = 1 - 7\alpha. \qquad (28)$$

This is due to the fact that such parameter values $\alpha$ correspond to low friction, when slip occurs throughout the contact ($\alpha \leq \alpha_{c1}$). As the value of $\alpha$ increases, the function $P(\gamma,\alpha)$ deviates from this straight line, as seen in the inset to the figure, which represents its enlarged fragment; it is clear from equation (26), that the point of first deviation from the straight line depends on the value of $\gamma$. After deviating from the linear behavior, all curves have a similar behaviour: first, the function $P(\gamma,\alpha)$ continues to decrease, and then, upon reaching the minimum, it increases (it is possible that the function has a local maximum) and converges towards a constant value corresponding to complete adherence, when $P(\gamma,\alpha)$ does not depend on $\alpha$ anymore and is described by the approximation

$$P(\gamma) \approx -1 + 2 \cdot \exp(-0.195\gamma) \cdot \cos^2[1.685(\gamma - 0.061)], \qquad (29)$$

obtained in [18]. The described behavior for large A is well traced in the three-dimensional diagram in Fig. 2. Also, Fig. 4b demonstrates, that the behavior for $\alpha = 12$ only insignificantly deviates from the solution of complete adherence (29), i.e. with an infinite coefficient of friction.

## 4 COMPARISON WITH THE EXPERIMENT

In many experiments devoted to the study of collision dynamics, a quantity called the coefficient of restitution is measured. It is the ratio of the velocities after and before the impact. Generally, the restitution coefficients of the normal and tangential velocity components are measured separately. We will consider the experiment carried out in [24], where the oblique elastic impact of a spherical body onto a plane was investigated. The coefficient of normal restitution for the elastic impact is trivially $e_n = 1$). The coefficient of restitution for the tangential motion depends on the collision conditions and is defined as

$$e_t = \frac{v_x}{v_{x0}}. \qquad (30)$$

In Fig. 5 the experimental results are from the work [24]. Each point in the figure was obtained by averaging at least 10 experiments with fixed values of the initial velocity and the angle of incidence. Moreover, in Fig. 5b the circles represent the direct measurements of the angular angular velocity, and the values indicated by the triangles are determined by the formula [24]

$$\omega = \frac{5}{2}\frac{v_0 \cos\alpha^\circ}{R}(1-e_t), \qquad (31)$$

based on the measurement results shown in Fig. 5a.

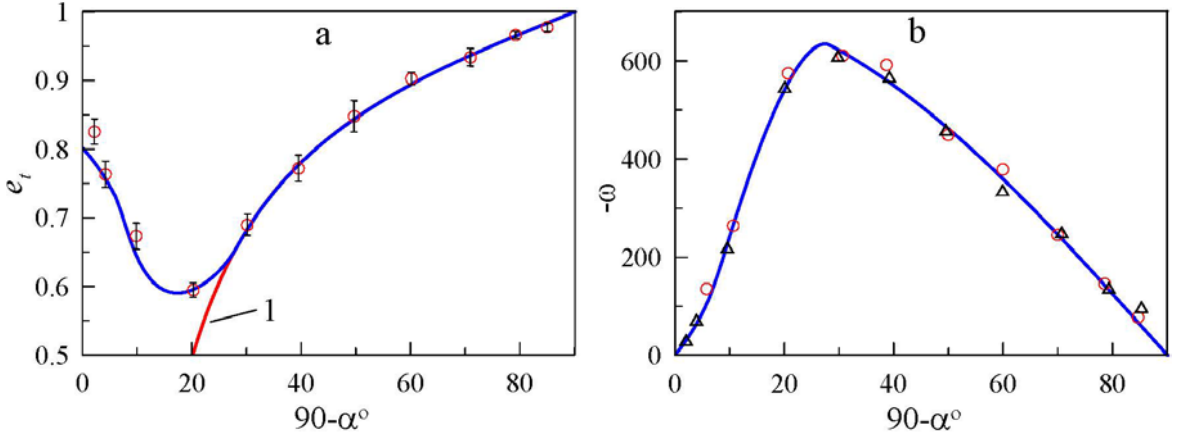

Fig. 5. The tangential recovery factor $e_t$ (30) (a) and the angular velocity $-\omega$ (b) as a function of incidence angle $(90-\alpha^\circ)$ in the region of elastic impact [24]. Curve 1 on the panel (a) is obtained from (35) with $\mu = 0.092$, $e_n = 1.0$. In both figures, the solid curves represent the simulation results within the framework of the MDR described above.

Since the data is obtained in the regime without plastic effects, we can compare these results with the results of the numerical experiments within the framework of the MDR described above. Namely, above we numerically calculated the function $P(\gamma,\alpha)$ (21), which allows us to determine all parameters of the collision. Moreover, the variable $\gamma$ (23) takes into account the parameters of the colliding bodies, and $\alpha$ (22) describes the influence of the friction coefficient $\mu$. In Fig. 5 the dependence of the restitution coefficient of the tangential velocity component $e_t$ (30) and the cyclic rebound velocity $\omega$, with zero initial rotation $\omega_0 = 0$, are given. Both can be calculated, if the function $P(\gamma,\alpha)$ is known. Using the relations (21), (1), (2) for $\omega_0 = 0$, we obtain:

$$e_t = \frac{2P(\gamma,\alpha)+5}{7}, \qquad (32)$$

$$\omega = \frac{5v_0 \cos\alpha^\circ \left[P(\gamma,\alpha)-1\right]}{7R}, \qquad (33)$$

where the radius of the sphere in the experimental setup was $R = 2.5 \cdot 10^{-3}$ m, and the initial velocity was $v_0 = 3.9$ m/s. The angle $\alpha°$ in formula (33) is the slip angle shown in Fig. 1. In [24] the angle of incidence was defined differently, which is $(90 - \alpha°)$ and plotted along the abscissa axis in Fig. 5. The value of $\alpha$ in the formulas (32), (33) depends on the slip angle $\alpha°$. Thus, in the case $\omega_0 = 0$ the parameter $\alpha$ is determined according to (22) as

$$\alpha(\omega_0 = 0) = \mu \tan \alpha°, \tag{34}$$

where the friction coefficient in the experimental setup is $\mu = 0.092$. Thus, using formulas (32), (33) and varying the angle $\alpha°$, it is possible to obtain the dependencies presented in Fig. 5 if the form of the function $P(\gamma, \alpha)$ shown in Fig. 2 and Fig. 4 is known. Before the beginning of the corresponding calculations it is necessary to determine the value of the parameter $\gamma$ (23), which remains constant for the series of experiments, the results of which are shown in Fig. 5. In [24] the parameters of the used materials are not given, but the materials themselves are indicated. Thus, sodium-calcium glass was used as the reflecting surface, and the reflected particle was made of aluminum oxide. There is no more detailed data in the work, and the given data is insufficient for the exact determination of the parameters, so we took typical values for these materials[1]. For the glass used we assumed $E_2 = 7.3 \cdot 10^{10}$ Pa and $\nu_2 = 0.217$; for aluminum oxide $E_1 = 3.7 \cdot 10^{11}$ Pa, $\nu_1 = 0.25$, it corresponds to a material containing 96% of $Al_2O_3$ at room temperature. This set of parameters allows us to find $E^*$ (5) and $G^*$ (6), which are equal to $E^* \approx 6.42 \cdot 10^{10}$ Pa, $G^* \approx 5.64 \cdot 10^{10}$ Pa, which according to the formula (23) leads to the value $\gamma \approx 1.754$.

The tangential component of the restitution coefficient $e_t$ and the cyclic frequency $\omega$ taken with the opposite sign are calculated from formulas (32) and (33) and are shown by solid lines in Fig. 5a and Fig. 5b, for $\gamma = 1.754$ and the parameter $\alpha$ according to the expression (34). The figure shows that the theoretical dependence obtained by us with great accuracy describes the experimental data throughout the whole range of the incidence angle. There is an undeniable advantage over the approximation

$$e_t = 1 - \mu(1 + e_n)\tan\alpha°, \tag{35}$$

---

[1]With parameter variation for different types of glasses and the percentage composition, the results obtained vary insignificantly. Note, that the value of $\gamma$ has not been subject to optimization to obtain best agreement between experiment and numerical model.

given in the work [13], and shown in Fig. 5a as curve 1, since it describes the behavior of the system only at incidence angles $(90° - \alpha) > 30°$. Once again we note that, since an elastic impact is considered, the normal coefficient of restitution is $e_n = 1.0$. At the same time, the results obtained by us and the approximation (35) ideally coincide in the domain of the latter ones applicability. It should be noted that in the case of an unknown nature of the contacting bodies, the results of particle collision experiments can be used to determine the elastic modulus of a material, for which it is possible to use relations (32), (33), since the form of the function $P(\gamma, \alpha)$ is known to us now. On the other hand, when using a sphere with specified parameters, it is possible to determine elastic parameters and the friction coefficient for modified surfaces which are often used in industry [25].

# 5 CONCLUSIONS

In the present paper we considered an experiment conducted by other authors on the collision of an elastic aluminum alloy particle with a glass surface. The dependence of the restitution coefficient of the tangential velocity component on the angle of incidence is studied in detail. The obtained theoretical results with high accuracy coincide with the experiments. To this date, in connection with the relevance of the problem under consideration in the study of granular dynamics, there is a large number of experimental studies, in which collisions of bodies made of various materials are studied. Using our results, we can accurately predict the behavior of colliding bodies in the experiment. However, it is worth noting that we only considered elastic impacts with a fixed coefficient of friction between the contacting surfaces. We did not take into account such effects as plastic deformation, adhesion forces between the contacting surfaces, surface roughness, the presence of inhomogeneities in bodies, the propagation of elastic waves, viscous friction in the air, the relaxation of elastic energy in a viscoelastic body. Hence, a complete picture of the dynamics of collisions is much more complicated, and there are still many unresolved issues. However, many of them can be solved within the dynamic contact modeling method described in the article.

## Acknowledgement

This work was supported in part by Tomsk State University Academic D.I. Mendeleev Fund Program and the Deutsche Forschungsgemeinschaft. I.A.L. is grateful to MESU for financial support under the Project No. 0116U006818.

## REFERENCES

[1] Ciamarra M. P., Lara A. H., Lee A. T. et al. Dynamics of drag and force distributions for projectile impact in a granular medium // Phys. Rev. Lett. 2004. V. 92, N 19. P. 194301.

[2] Jop P., Forterre Y., Pouliquen O. A constitutive law for dense granular flows // Nature. 2006. V. 441, N 7094. P. 727-730.

[3] Bernard B. Impacts in mechanical systems: analysis and modelling / B. Bernard. Berlin, New York: Springer, 2000.

[4] Parteli E. J. R., Schmidt J., Blümel C. et al. Attractive particle interaction forces and packing density of fine glass powders // Sci. Rep. 2014. V. 4. P. 6227 (7 pp.).

[5] Maw N., Barber J. R., Fawcett J. N. The oblique impact of elastic spheres // Wear. 1976. V. 38, N 1. P. 101-114.

[6] Hauger W. Technische Mechanic. Bd 3: Kinetik 7 / W. Hauger, W. Schnell, D. Gross. Berlin: Springer, 2002.

[7] Hertz H. J. Ueber die Berührung fester elastischer Körper // J. für die reine Angew. Math. 1882. V. 92. P. 156-171.

[8] Mindlin R. D. Compliance of elastic bodies in contact // ASME J. Applied Mech. 1949. V. 16. P. 259-268.

[9] Barber J. R. Adhesive contact during the oblique impact of elastic spheres // Appl. Math. Phys. (ZAMP). 1979. V. 30, N 3. P. 468-476.

[10] Maw N., Barber J. R., Fawcett J. N. The role of elastic tangential compliance in oblique impact // Trans. ASME: J. Lubr. Technol. 1981. V. 103, N 1. P. 74-80.

[11] Labous L., Rosato A. D., Dave R. N. Measurements of collisional properties of spheres using high-speed video analysis // Phys. Rev. E. 1997. V. 56, N 5. P. 5717-5725.

[12] Foerster S. F., Louge M. Y., Chang H. et al. Measurements of the collision properties of small spheres // Phys. Fluids. 1994. V. 6, N 3. P. 1108-1115.

[13] Sondergaard R., Chaney K., Brennen C. E. Measurements of solid spheres bouncing off flat plates // ASME J. Appl. Mech. 1990. V. 57, N 3. P. 694-699.

[14] Stronge W. J. Impact Mechanics / W. J. Stronge. Cambridge: Cambridge University Press, 2004.

[15] Jia Y. -B. Three-dimensional impact: energy-based modeling of tangential compliance // The International Journal of Robotics Research. 2013. V. 32, N 1. P. 56-83.

[16] Popov V. L., Psakhie S. G., Numerical simulation methods in tribology: possibilities and limitations // Tribology International. 2007. V. 40, N. 6. P. 916-923.

[17] Popov V. L. Method of dimensionality reduction in contact mechanics and friction / V. L. Popov, M. Heß. Berlin: Springer, 2014.

[18] Lyashenko I. A., Popov V. L. Impact of an elastic sphere with an elastic half space revisited: Numerical analysis based on the method of dimensionality reduction // Sci. Rep. 2015. V. 5. P. 8479 (5 pp.).

[19] Willert E., Popov V. L. Impact of an elastic sphere with an elastic half space with a constant coefficient of friction: Numerical analysis based on the method of dimensionality reduction // ZAMM - Journal of Applied Mathematics and Mechanics / Zeitschrift für Angewandte Mathematik und Mechanik. 2016. DOI: 10.1002/zamm.201400309.

[20] Lyashenko I. A., Willert E., Popov V. L. Adhesive impact of an elastic sphere with an elastic half space: Numerical analysis based on the method of dimensionality reduction // Mech. Mater. 2016. V. 92. P. 155-163.

[21] Cattaneo C. Sul contatto di due corpi elastici: distribuzione locale degli sforzi // Rendiconti dell'Accademia nazionale dei Lincei. 1938. V. 27, N 6. P. 342-348; 434-436; 474-478.

[22] Munisamy R. L., Hills D. A., Nowell D. Static axisymmetric hertzian contacts subject to shearing forces // ASME J. Appl. Mech. 1994. V. 61, N 2. P. 278-283.

[23] Landau L. D. Theory of elasticity / L. D. Landau, E. M. Lifshitz. Oxford: Pergamon Press, 1986.

[24] Gorham D. A., Kharaz A. H. The measurement of particle rebound characteristics // Powder Technol. 2000. V. 112, N 3. P. 193-202.

[25] A. D. Pogrebnjak, S. N. Bratushka, V. M. Beresnev, and N. Levintant-Zayonts, Russ. Chem. Rev. **82**, 1135 (2013).